\documentclass{article}
\usepackage{graphicx} 
\usepackage{authblk}  
\usepackage{amssymb}
\usepackage{amsmath}
\usepackage{subcaption}
\usepackage{adjustbox}
\usepackage{array}
\usepackage{multirow}
\usepackage{adjustbox}
\usepackage{makecell}
\usepackage{xcolor}

\title{Comparative Analysis of Plasticity-based GND Density Estimation
Methods in Crystal Plasticity Finite Element Models}

\author[1]{Michael Pilipchuk}
\author[2]{Chaitali Patil}
\author[1,*]{Veera Sundararaghavan}

\affil[1]{Department of Aerospace Engineering, University of Michigan}
\affil[2]{Department of Materials Science and Engineering}
\affil[*]{Corresponding author. Email: \texttt{veeras@umich.edu}}

\date{}

\begin{document}

\maketitle

\begin{abstract}
In crystal plasticity finite element (CPFE) simulations, accurately quantifying geometrically necessary dislocations (GNDs) is critical for capturing strain gradients in polycrystals. We compare different methods for quantifying GNDs, all of which originate from the Nye tensor, which is computed as the curl of the plastic deformation gradient. The projection technique directly decomposes the Nye tensor onto individual screw and edge dislocation components to compute GNDs. This approach requires converting a nine-component Nye tensor into densities for a larger number of dislocation systems, a fundamentally underdetermined (non-unique) process, which is resolved using $L2$ minimization. In contrast, when employing CPFE analysis, one could directly compute dislocation densities on each slip system using shear gradients. Projection and slip gradient methods are compared with respect to their prediction of GNDs with changing grain size, strain, and grain neighborhoods, including multigrain junctions. Although these techniques match analytical GND densities for single slip, single crystal deformation, and are consistent with anticipated overall GND trends, we find that the GND densities from projection techniques are significantly lower than those predicted from CPFE-based slip gradients in polycrystals. A suggested improvement of only using the active dislocation systems in the projection technique almost entirely resolved this mismatch.
\end{abstract}

\section{Introduction}
Geometrically necessary dislocation (GND) density can be directly correlated with the gradients in the deformation during inhomogeneous plastic deformation~\cite{Ashby1970,gao2003}. As a result, quantification of GNDs has become fundamental in understanding the size-dependent plastic behavior of polycrystalline metals~\cite{fleck1994}. Experimentally, the presence of GNDs has been linked to phenomena such as indentation size effects~\cite{nix1998} and the grain size effect (Hall–Petch relationship), where the strengthening of metals scales inversely with grain size~\cite{hall1951,armstrong1962}. In micro-indentation tests, for instance, the local strain gradients generated beneath the indenter tip lead to GND accumulation~\cite{wilkinson2010,mattucci2021}, which in turn influences hardness and yield strength~\cite{nix1998,durst2006,javaid2018}. Similarly, the Hall–Petch effect, where finer grains yield higher strengths, is interpreted in terms of enhanced GND densities at grain boundaries that accommodate lattice curvature~\cite{kuhlmann1991}.

At the mechanistic level, GNDs originate from the incompatibility inherent in the multiplicative decomposition of the total deformation gradient. In crystal plasticity, the total deformation gradient \( \mathbf{F} \) is decomposed into an elastic part \( \mathbf{F}^e \) and a plastic part \( \mathbf{F}^p \) as $
\mathbf{F} = \mathbf{F}^e \mathbf{F}^p$. While \(\mathbf{F}\) must satisfy compatibility (\(\text{curl}(\mathbf{F}) = \mathbf{0}\)), neither \(\mathbf{F}^e\) nor \(\mathbf{F}^p\) is curl-free. The Nye tensor, \(\mathbf{G} = \text{curl}(\mathbf{F}^p)\), first introduced by Kröner~\cite{Kroener1958}, directly quantifies GND density by encoding the Burgers vector content per unit volume required to accommodate plastic strain gradients. In polycrystals, these gradients naturally arise at grain boundaries and multigrain junctions due to heterogeneous slip activation, with GND densities scaling inversely with grain size as predicted by Ashby~\cite{Ashby1970}.  Gradient-based density calculations are now commonly used for estimation of the dislocation density accumulation with advanced characterization techniques like high resolution electron back-scattered diffraction microscopy (HR-EBSD)~\cite{jiang2015,kalacska2020,sulzer2020,witzen2023,zhong2024,peng2025}, TriBeam tomography~\cite{witzen2024} or high energy diffraction microscopy (HEDM)~\cite{gustafson2023,li2024} indirectly through elastic lattice rotations, \(\mathbf{R}^e\)~\cite{field2005,pantleon2008}, derived from \(\mathbf{F}^e\). Similarly, complementary CPFE analysis is being used with a range of advanced GND estimation methods~\cite{Cermelli2001,Arsenlis1999,Dai1997} to corroborate experimental observations~\cite{dunne2012,wan2016,bandyopadhyay2021} or to understand the mechanistic origins of important physical phenomena~\cite{mackey2023}. Hence, it is imperative to understand the inherent similarities or differences in the computational methods for GND estimation. 

A critical aspect of this tensorial measure is that \(\mathbf{G}\)’s nine independent components cannot uniquely determine the dislocation densities associated with the multiple slip systems present in face-centered cubic (FCC) metals. This underdetermination stems from the fact that a single \(\mathbf{G}\) configuration can correspond to numerous dislocation arrangements: edge dislocations along the \(\langle112\rangle\) directions and screw dislocations along the \(\langle 110 \rangle\) directions; both contribute to \(\mathbf{G}\) but require separate projections onto slip systems. 

In this work, three different methods are used to resolve the underdetermination of \(\mathbf{G}\): First, the pseudoinverse decomposition minimizes the L2-norm of dislocation densities, distributing \(\mathbf{G}\)’s components across dislocation systems via the Moore-Penrose pseudoinverse~\cite{Arsenlis1999}. The second is based on recent works suggesting limiting this projection to the set of active slip systems instead of the complete set~\cite{Demir2024}. Finally, the CPFE-based method directly employs gradients of the slip strain \(\gamma^\alpha\), to uniquely determine GND densities on each slip system~\cite{Dai1997,Ma2006}.   

Our simulations demonstrate that these methods show results similar to those predicted analytically and that with mesh refinement, GND densities converge and exhibit the expected experimental trends in all these methods: (i) an increase in GND density with decreasing grain size, (ii) a rise in GND density with increasing applied strain, in agreement with classical models. Interestingly, we find that the magnitudes of GNDs found by projection onto all dislocation systems and CPFE-based methods are significantly different in polycrystals. The projection method, which relies on a pseudoinverse to distribute a nine-component Nye tensor over several dislocation systems, tends to consistently provide lower dislocation densities compared to the slip gradient method, a discrepancy which is almost entirely resolved by limiting the projection to the active dislocation systems. These methods are compared with respect to cases of changing grain size and strain. Furthermore, the effect of grain neighborhoods on GND densities is also investigated. 

\section{Methodology}

In crystal plasticity theory, a plastic flow rule is defined for the rate of the plastic deformation gradient, $\dot{\mathbf{F}}^p$, in terms of the plastic shear rate on slip system $\alpha$, $\dot{\gamma}^\alpha$, and the slip system's undeformed direction and normal, $\mathbf{m}_0^\alpha$ and $\mathbf{n}_0^\alpha$, respectively.

\begin{equation}  
\dot{\mathbf{F}}^p{\mathbf{F}^p}^{-1} = \sum_\alpha {\dot{\gamma}^\alpha \mathbf{m}_0^\alpha \otimes \mathbf{n}_0^\alpha}
\end{equation}

Using Euler forward difference based on the plastic deformation gradient in the previous time step $\mathbf{F}^p_n$ and change in slip shear $\Delta{\gamma}^\alpha=\dot{\gamma}^\alpha \Delta t$ over time step $\Delta t$, this is numerically written as:\\

\begin{equation} 
    (\mathbf{F}^p - \mathbf{F}^p_n){\mathbf{F}^p_n}^{-1} = \sum_\alpha {\Delta{\gamma}^\alpha \mathbf{m}_0^\alpha \otimes \mathbf{n}_0^\alpha}
\end{equation}

\begin{equation}
    \mathbf{F}^p   = (\mathbf{I} + \sum_\alpha {\Delta{\gamma}^\alpha \mathbf{m}_0^\alpha \otimes \mathbf{n}_0^\alpha})\mathbf{F}^p_n
\end{equation}

For the present study, the analysis was restricted to the small deformation regime (keeping linear terms in $\Delta \gamma^\alpha$) to study metals undergoing modest plastic strains typically on the order of 5–10\%, leading to the simplified expression:

\begin{equation}\label{eq:shearSchmidInt}
    {\mathbf{F}}^p = \mathbf{I} +\sum_\alpha {{\gamma}^\alpha \mathbf{m}_0^\alpha \otimes \mathbf{n}_0^\alpha}
\end{equation}

where $\gamma^\alpha = \sum_n \Delta \gamma^\alpha$, the total slip strain on system $\alpha$ summing $n$ increments of slip strain.

\subsection{Crystal plasticity finite element preliminaries} \label{subsec:cpfePrelim}

The derivation of the Nye tensor~\cite{Nye1953} begins from the multiplicative decomposition of the deformation gradient~\cite{Yaghoobi2019}. In addition to the plastic deformation gradient defined above, the transformation from an undeformed state $\mathbf{X}$ to the final state $\mathbf{x}$ using a deformation gradient tensor $\mathbf{F}$ is modeled using multiplicative Kröner-Lee decomposition~\cite{Kroner1959,Lee1969} using an elastic deformation gradient ($\mathbf{F}^e$) such that:

\begin{equation} \label{eq:Fdeco}
    \mathbf{F} = \mathbf{F}^e \mathbf{F}^p   
\end{equation}

\begin{figure}
    \centering
    \includegraphics[width=.65\linewidth]{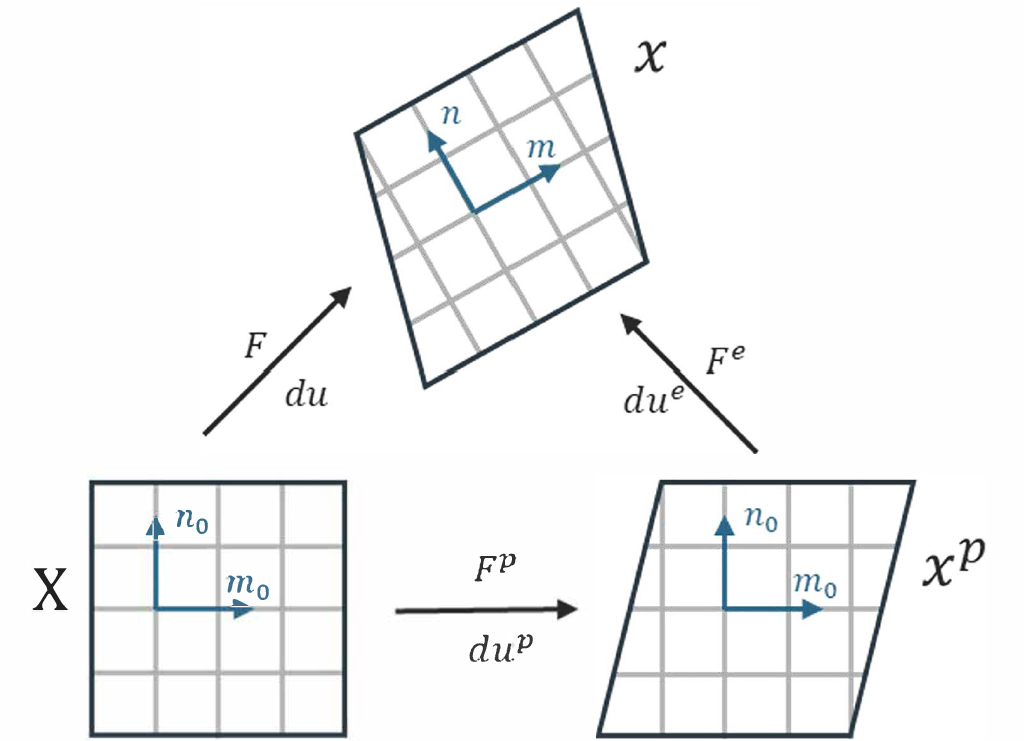}
    \caption{Deformation gradient decomposition.}
    \label{fig:Fig01-gradDef}
\end{figure}

An intermediate state, $\mathbf{x}^p$, accommodates the plastic deformation, as shown in Figure~\ref{fig:Fig01-gradDef}. An infinitesimally small line segment, $d\mathbf{X}$, in the undeformed state, is related to $d\mathbf{x}$ in the deformed state by: 

\begin{equation}
d\mathbf{x} = \mathbf{F}d\mathbf{X} =\mathbf{F}^e\mathbf{F}^pd\mathbf{X}   
\end{equation}

This transformation is also captured using the difference between the two states.

\begin{equation} \label{eq:up}
    d\mathbf{u}^p = d\mathbf{x}^p - d\mathbf{X} = \mathbf{F}^pd\mathbf{X} - d\mathbf{X} = (\mathbf{F}^p-\mathbf{I})d\mathbf{X}
\end{equation}

Assuming the isochoric nature of plastic deformation, $J_p=1$. The conversion from an element in the reference configuration, $\mathbf{n}_RdA_R$, to the deformed plastic configuration, $\mathbf{n}dA$, is simplified.
\begin{equation} \label{eq:ref2lat}
    \mathbf{n}dA = J_p{\mathbf{F}^p}^{-T}\mathbf{n}_RdA_R={\mathbf{F}^p}^{-T}\mathbf{n}_RdA_R
\end{equation}

\subsection{Geometric dislocation tensor} \label{subsec:nyeTensor}
Nye~\cite{Nye1953} related the vector $\mathbf{B}$ needed to close a Burgers circuit of unit area with normal $\mathbf{n}$ to the Nye tensor, $\mathbf{G}$, assuming that the dislocations involved are continuously distributed.

\begin{equation}
    \mathbf{B} = \mathbf{G}\mathbf{n}
\end{equation}

The vector $\mathbf{B}$ is the sum of all the Burgers vectors, $\mathbf{b}$ acting in that direction, such that $\mathbf{B} = n_d \mathbf{b}$, where $n_d$ is the number of dislocations.
Since $\mathbf{B}$ describes the vector needed to close the circuit, it can be found from the circuit integral of plastic deformation. Then, for a circuit of arbitrary area,

\begin{equation} \label{eq:circuitclose}
    \mathbf{B} = \iint \mathbf{G}\mathbf{n}dA = \oint \,d\mathbf{u}^p
\end{equation}

Substituting the plastic displacement from Eq.~\ref{eq:up}:

\begin{equation} \label{eq:compatibilityIntegral}
    \oint \,d\mathbf{u}^p = \oint \,(\mathbf{F}^p-\mathbf{I})d\mathbf{X}
\end{equation}

Using Stokes' theorem, the line integral becomes a surface integral of the curl \footnote{Das et al.~\cite{Das2018} concisely lists the various definitions of curl used in GND literature. Here, the curl is defined such that:

\begin{equation}
    (\text{curl}\mathbf{A})_{ij} = (-\nabla \times \mathbf{A})_{ij} = \varepsilon_{jmn}A_{in,m} \nonumber
\end{equation}

}.

\begin{equation} \label{eq:stokes}
    \oint \,(\mathbf{F}^p-I)dX = \iint \, \text{curl}(\mathbf{F}^p-\mathbf{I})\mathbf{n}_RdA_R= \iint \, \text{curl}(\mathbf{F}^p)\mathbf{n}_RdA_R
\end{equation}

Using the change between the current and reference configurations in Eq.~\ref{eq:ref2lat}, the substitution of Eqs.~\ref{eq:compatibilityIntegral} and \ref{eq:stokes} into Eq.~\ref{eq:circuitclose} yields:
\begin{equation}
    \int \,\mathbf{G}\mathbf{n}dA = \int \, \text{curl}(\mathbf{F}^p){\mathbf{F}^p}^T\mathbf{n}dA
\end{equation}

Finally, the geometric dislocation tensor is:
\begin{equation} \label{eq:curlFp}
    \mathbf{G} =\text{curl}(\mathbf{F}^p){\mathbf{F}^p}^T \text{,}
\end{equation}

which, for small deformations, can be simplified, leading to the small deformation formulation in Eq.~\ref{eq:curlFpSimp}.

\begin{equation} \label{eq:curlFpSimp}
    \mathbf{G} =\text{curl}(\mathbf{F}^p)
\end{equation}

Unlike CPFE, experimental approaches tend to compute GND densities from lattice curvature, such as in Ref.~\cite{Britton2011}. There are equivalent definitions for the Nye tensor using both the plastic and elastic deformation gradient tensors (based on the multiplicative decomposition in Eq.~\ref{eq:Fdeco} and the fact that $\text{curl}(\mathbf{F}) = 0$ for compatibility. Hence, as shown in Ref.~\cite{Cermelli2001}, the relationship in Eq.~\ref{eq:FeFpRelate} can be found, written in the full form without the simplifying assumptions.

\begin{equation} \label{eq:FeFpRelate}
    \mathbf{G} = {J^p}^{-1}\text{curl}(\mathbf{F}^p){\mathbf{F}^p}^T ={J^e}\text{curl}({\mathbf{F}^e}^{-1}){\mathbf{F}^e}^{-T} 
\end{equation}

The elastic deformation gradient tensor can be decomposed into the elastic stretch, $\mathbf{U}^e$, and rotation $\mathbf{R}^e$.
\begin{equation}
    \mathbf{F}^e = \mathbf{R}^e\mathbf{U}^e
\end{equation}

By assuming that the elastic stretch contribution is small, $\mathbf{U}^e \approx \mathbf{I}$~\cite{wilkinson2010,jiang2015}, an estimate of Nye tensor based on the measured lattice rotations is also used in practice:
\begin{equation} \label{eq:expNye}
     \mathbf{G} \approx \text{curl}({\mathbf{R}^e}^T)
\end{equation}
Due to the equivalence in Eq. \ref{eq:FeFpRelate} and at small strains, the trends for dislocation density computed from the plastic deformation gradient should be similar to those measured in experiments. 

\subsection{GND density: projection onto dislocation systems}\label{subsec:disSys}
In the Nye tensor formulations~\cite{Nye1953}, the geometric dislocation tensor is defined as the number of dislocations, $n_d$, with Burgers vector $\mathbf{b}$, crossing a unit area with a unit normal $\mathbf{r}$:

\begin{equation}
    \mathbf{G} = n_d \mathbf{b} \otimes \mathbf{r} 
\end{equation}

Arsenlis and Parks~\cite{Arsenlis1999} then decomposed the dislocation tensor, defining it as the product of GND densities ($\rho_{GND}^\alpha$) with their dyadic ($\mathbf{d}^\alpha$) summed over all unique screw and edge dislocation systems, listed in Arsenlis and Parks~\cite{Arsenlis1999}.

\begin{equation} \label{eq:sumForm}
    \mathbf{G} = \sum_{\alpha=1}^{N} \rho_{GND}^\alpha \mathbf{d}^\alpha 
\end{equation}

The dyadic is defined as the outer product of the Burgers vector and $\mathbf{t}$, a unit-length tangent vector:
\begin{equation}
    \mathbf{d}^\alpha = \mathbf{b}^\alpha \otimes \mathbf{t}^\alpha
\end{equation}

Rewriting the Nye tensor as a column vector, $\mathbf{\Lambda}$, and organizing the GND densities on each plane into a column vector, $\mathbf{\rho}_{GND}$, results in the underdetermined system in Eq.~\ref{eq:rhoGNDG}~\cite{Arsenlis1999}. The operator $\mathbf{A}$ is found from the dyadic containing the Burgers vector and the tangent vector. 

\begin{equation}\label{eq:rhoGNDG}
    \mathbf{\Lambda} = \mathbf{A}\mathbf{\rho}_{GND}
\end{equation}

The underdetermined system can be solved with either $L1$ or $L2$ minimization. For $L1$, the total dislocation length is minimized under the assumption that this is the lowest energy configuration~\cite{Demir2023}. $L2$ is a least-squares approach~\cite{Demir2023} that minimizes the sum of squares of the GND density on each edge and screw system, $\sqrt{\sum {\rho^\alpha_{GND}}^2}$~\cite{Arsenlis1999}.
Although the maximum GND density observed in a grain can be up to an order of magnitude less using $L1$ than $L2$~\cite{Demir2023}, the total dislocation densities have been found to be similar~\cite{Das2018}. Thus, for finding the total density, the $L2$ approach provides similar results while using a simpler formulation~\cite{Arsenlis1999}, and so it is used here.

Equation \ref{eq:rhoGNDG} is then solved by $L2$ minimization based on Morse-Penrose pseudoinverse, using Eq.~\ref{eq:L2Minimization} to project the Nye tensor onto dislocation systems.

\begin{equation}
    \label{eq:L2Minimization}
    \mathbf{\rho}_{GND} = \mathbf{A}^T(\mathbf{A}\mathbf{A}^T)^{-1}\mathbf{\Lambda}
\end{equation}

\subsection{GND density: active dislocation set approach}
The previous approach uses all dislocation systems and does not account for the fact that some dislocation systems may be inactive depending on the deformation mode. In Demir et al.~\cite{Demir2024}, a solution is proposed to this by restricting GND calculation to only active dislocation systems that have cleared a threshold amount of slip activity. Systems with $\gamma \leq 1\times 10^{-8}$ are removed. The non-zero value is selected to account for numerical errors. 

The remainder of the derivation is similar to the previous section, but with a variable number of active dislocation systems. As before, Eq.~\ref{eq:sumForm} relates the Nye tensor to the GND density and dyadic for each system, but this time only the active ones. Then, at each integration point in the finite element mesh, the dimensions of $\mathbf{A}$ in Eq.~\ref{eq:rhoGNDG} are variable.

\subsection{GND density: shear approach}

The third approach to determining GND densities is based on the approach in Ref.~\cite{Dai1997}. Substituting the definition of the plastic deformation tensor from Eq.~\ref{eq:shearSchmidInt} into the one for the Nye tensor in Eq.~\ref{eq:curlFpSimp}, the equation becomes

\begin{equation} \label{eq:nyeShear}
    \mathbf{G} = \text{curl}(\mathbf{F}^p) = -\nabla\times\sum_\alpha \gamma^\alpha \mathbf{m}_0^\alpha \otimes \mathbf{n}_0^\alpha \text{ .}
\end{equation}

The total Nye tensor is a sum of the contributions from every slip system:
\begin{equation}
    \mathbf{G} = \sum_\alpha \mathbf{G}^\alpha \text{ ,}
\end{equation}

which means that for each slip system $\alpha$, the Nye tensor is:

\begin{equation} \label{eq:baseNyeAlpha}
    \mathbf{G}^\alpha = -\nabla \times (\gamma^\alpha \mathbf{m}_0^\alpha \otimes \mathbf{n}_0^\alpha) \text{ .}
\end{equation}

For small deformations, the slip direction and normal can be considered constant. A new vector, defined as the normal to the GNDs on a slip system and parallel to their tangent vector, $\mathbf{n}_{GND}^\alpha$, is in Eq.~\ref {eq:nyeGNDNorm}~\cite{Ma2006}

\begin{equation} \label{eq:nyeGNDNorm}
    \mathbf{G}^\alpha = \rho_{GND}^\alpha b^\alpha \mathbf{m}_0^\alpha\mathbf{n}_{GND}^\alpha
\end{equation}

Equations~\ref{eq:baseNyeAlpha} and~\ref{eq:nyeGNDNorm} are then combined.

\begin{equation}
    \rho_{GND}^\alpha b^\alpha \mathbf{m}_0^\alpha\mathbf{n}_{GND}^\alpha = -\nabla \times (\gamma^\alpha \mathbf{m}_0^\alpha \otimes \mathbf{n}_0^\alpha) 
\end{equation}

Solving for the GND density on each slip system and taking the magnitude gives:

\begin{equation}\label{eq:shearGNDperAlpha}
    \rho_{GND}^\alpha = \frac{1}{b^\alpha} ||\nabla\gamma^\alpha \times \mathbf{n}_0^\alpha|| \text{ ,}
\end{equation}

Finally, the summation can be returned to get the total GND state, in Eq.~\ref{eq:shearFinal} \footnote{The large deformation formulation uses the rate form of the equations~\cite{Ma2006,Bargmann2012}. The rate of GND density accumulation, $\dot{\rho}_{GND}^\alpha$ based on the shear rate, $\dot{\gamma}^\alpha$  described below need to be integrated over time:
\begin{equation*}
    \dot{\rho}_{GND}^\alpha = \frac{1}{b^\alpha}||\nabla\times(\dot{\gamma}^\alpha \mathbf{F}^p)\mathbf{n}_0^\alpha||
\end{equation*}
}.

\begin{equation} \label{eq:shearFinal}
        \rho_{GND} = \sum_\alpha \frac{1}{b^\alpha} ||\nabla\gamma^\alpha \times \mathbf{n}_0^\alpha||
\end{equation}

\subsection{Test models} \label{subsec:scAnalysis}
The above approaches were implemented as post-processing tools for PRISMS-Plasticity, an open-source CPFE package~\cite{Yaghoobi2019}. To test them, a series of simulations was run using three different models. All three were based on copper, as calibrated in Ref.~\cite{Yaghoobi2019}.

The first set of simulations uses two cubic models with 1 mm sides, discretized with $15^3$ elements. They are single crystals oriented along [000]. Two loading conditions were imposed: symmetrical uniaxial tension, representing a gradient-free case, and a case with gradients using the loading conditions adapted from Demir et al.\cite{Demir2024}. One face is fixed, while the opposite is loaded in shear, increasing from $\gamma=0$ on one edge up to $\gamma=0.05$.

\begin{figure}
    \centering
    \includegraphics[width=1\linewidth]{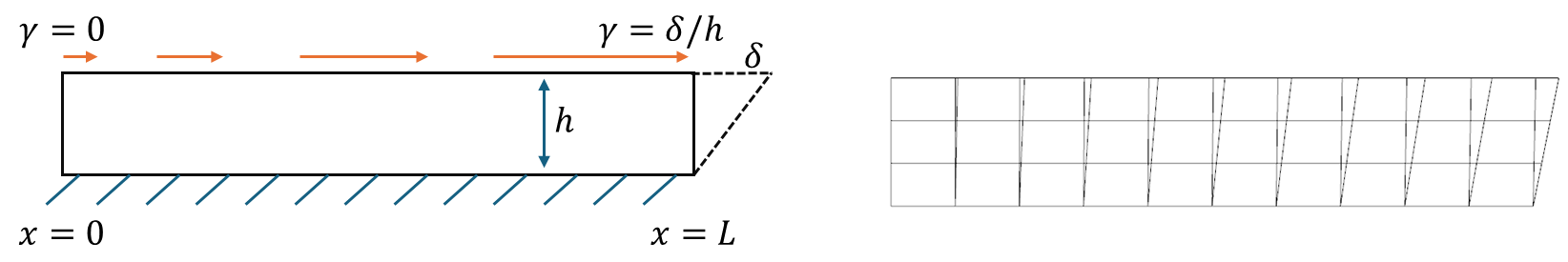}
    \caption{Single crystal beam shear loading and dimensioning adapted from~\cite{Demir2024} (left) and mesh in the undeformed and deformed configurations (right).}
    \label{fig:Fig02-scDiagram}
\end{figure}

The second single crystal case, adapted from~\cite{Demir2024}, is also a single crystal along [000], shaped as a beam ($5 \text{mm} \times 1 \text{ mm}^2$), discretized by 10$\times$3$\times$3 elements. This uses the same loading condition as the model above: one long face is fixed and the opposing face is loaded in shear, increasing from zero to 1\% strain along the length of the beam, illustrated in Fig~\ref{fig:Fig02-scDiagram}. For this simulation, the standard FCC slip systems are replaced by a single slip system with direction $\mathbf{m} = [100]$ and normal $\mathbf{n} = [001]$ to accommodate the deformation.

In the single slip case, the edge dislocation density can be analytically predicted from $\rho_{GND} = -\frac{1}{b}\nabla\gamma\cdot\mathbf{m}$~\cite{Arsenlis1999}. For this beam, the analytical solution predicts $\rho_{GND} = 7812.5$ mm$^{-2}$.

\begin{figure}
    \centering
    \includegraphics[width=0.35\linewidth]{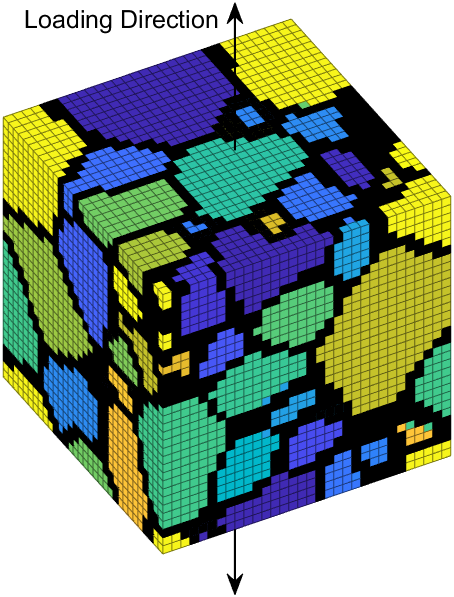}
    \caption{Sample microstructure and loading direction used for GND analysis.}
    \label{fig:Fig03-microstructure}
\end{figure}

The remaining work is done with a polycrystalline RVE using copper calibrated by~\cite{Yaghoobi2019}. A synthetic test microstructure was generated in DREAM.3D~\cite{Groeber2014} and shown in Fig.~\ref{fig:Fig03-microstructure}. 
It is a cube with 5 $\mu m$ sides and includes 64 grains with an average equivalent sphere diameter (ESD) of 1.17 $\mu m$.
The sample is modeled with a $34^3$ mesh and loaded in tension to 10\% strain using PRISMS-Plasticity's rate-independent model, along the direction indicated in Fig.~\ref{fig:Fig03-microstructure}. GNDs were computed at 18 strain levels during the loading. Grain boundaries (GBs) are not separately modeled and are the interface between elements of adjacent grains. For visualization purposes, they have been overlain on the polycrystal simulation results to highlight behaviors relative to the GB.

\section{Results and Discussion}

\subsection{Single crystal simulations}
\begin{figure}
    \centering
    \includegraphics[width=0.7\linewidth]{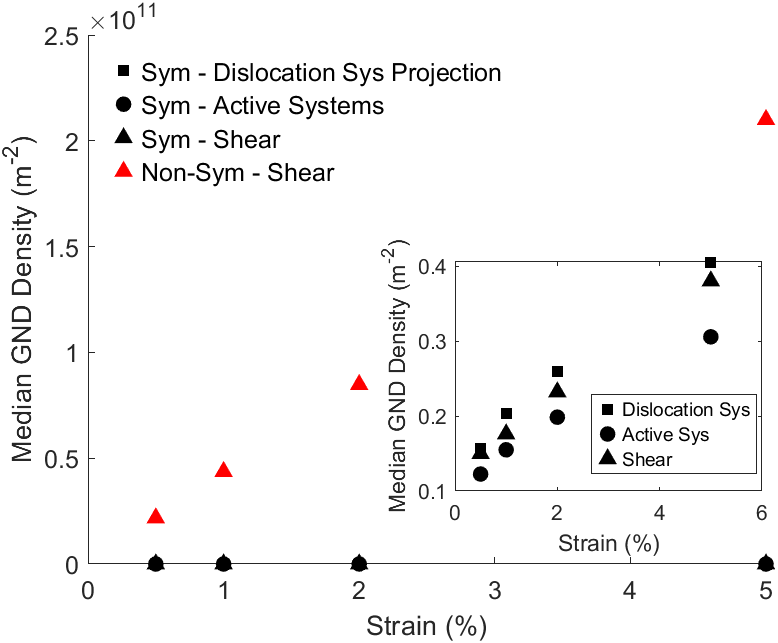}
    \caption{Median GND density in the symmetric tension vs non-symmetric shear case. The inset shows an adjusted GND density axis for clarity with the symmetric case.}
    \label{fig:Fig04-scSymNonSym}
\end{figure}

The symmetrically loaded single crystal case is theoretically gradient-free, leading to zero dislocation density. The non-uniform shear case is expected to have gradients, and thus non-zero dislocation densities. In Figure~\ref{fig:Fig04-scSymNonSym}, the median GND density for both models is plotted against the strain. As shown in the inset, there is some noise observed for the symmetric cases, but it is negligible compared to the densities observed for the non-symmetric case, which only has the shear approach plotted for comparison.

\begin{figure}
    \centering
    \includegraphics[width=1\linewidth]{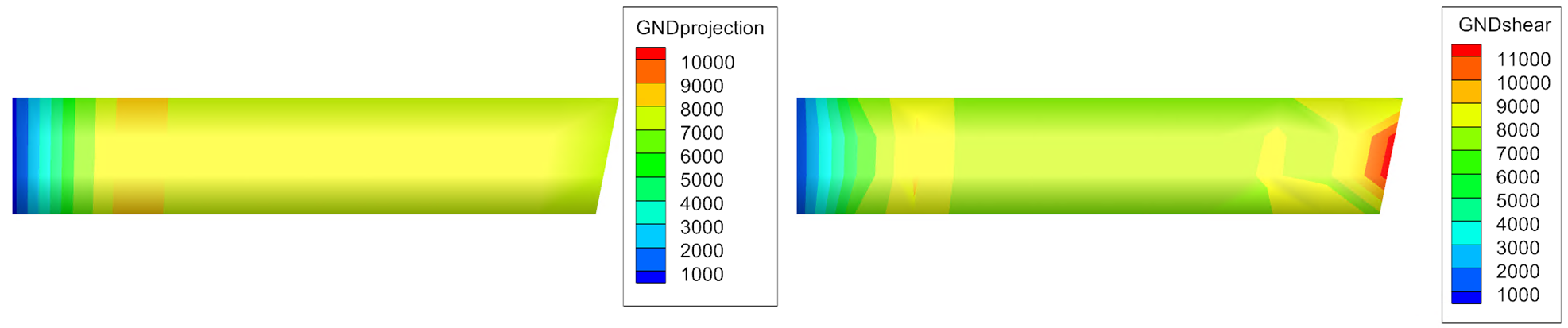}
    \caption{GND density distributions for the sheared single crystal beam using (a) the dislocation system approach and (b) the shear approach.}
    \label{fig:Fig05-scShearGND}
\end{figure}

GND density distributions for the sheared beam are in Figure~\ref{fig:Fig05-scShearGND} (in mm$^{-2}$). As previously stated, for this beam, the analytically-derived GND density is $\rho_{GND} = 7812.5$ mm$^{-2}$. The median GND density using the dislocation system approach is 7773.2 mm$^{-2}$ and for the shear approach it is 7779.7~mm$^{-2}$, 0.5\% and 0.42\% error, respectively. 
Error from the analytical solution is greatest at the simulation volume boundaries because it involves finding gradients, which in turn require neighboring nodal values. While the use of periodic boundary conditions can alleviate this effect, it is not always possible to implement them, such as for this simulation. Considering the model without the removal of the boundary elements, as shown in Fig. \ref{fig:Fig05-scShearGND}, the mean GND density is 7207.4 mm$^{-2}$ for the shear approach. However, if the boundary elements are removed, the average is 7368.4 mm$^{-2}$. Thus, keeping the boundary in the model would theoretically result in a less accurate prediction of the total average GND density. One aspect to note is that the GNDs are overpredicted on one side and underpredicted on the other side, so while the GND densities of grains near the RVE edges are expected to be different, it can either underpredict or overpredict based on the location. For all the polycrystal results in the manuscript,  this external layer in the RVE was removed to avoid biasing the results.

\begin{figure}
    \centering
    \includegraphics[width=0.7\linewidth]{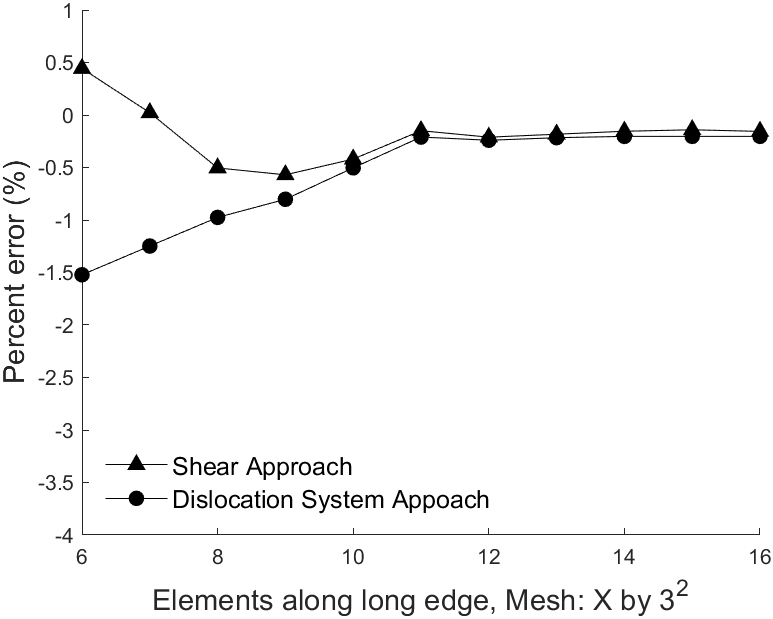}
    \caption{Mesh convergence for the single slip single crystal sheared beam studying mesh refinement along the long side of the beam.}
    \label{fig:Fig06-scShearGNDStudies}
\end{figure}

Since the GND densities are computed in a discretized mesh, the effects of discretization were observed by coarsening and refining the long edge in the range of 6 to 16 elements. 
In Fig.~\ref{fig:Fig06-scShearGNDStudies}, the resulting error is plotted against the number of elements used for the long edge. 
It is noteworthy that at all tested values, the shear approach gives higher GND densities, which are closer to the analytical model than the dislocation system approach. As there is only one dislocation system used, the active system approach would give the same results as the dislocation approach.

\subsection{GND density distributions}
For the studies with the polycrystal RVE, the nodal values were averaged at the element centers. Then, the outermost element layer on every side was removed to mitigate the effects of boundary conditions in computing the slip gradients. 

\begin{figure}
\centering
\scalebox{1}{
\begin{tabular}{
    >{\arraybackslash}m{0.16\textwidth}  
    >{\centering\arraybackslash}m{0.27\textwidth}  
    >{\centering\arraybackslash}m{0.27\textwidth}  
    >{\centering\arraybackslash}m{0.27\textwidth}  
}

    \textbf{Strain:} & \textbf{1.5\%} & \textbf{5\%} & \textbf{10\%} \\ 

    \makecell[l]{\textbf{Dislocation} \\ \textbf{System}\\ \textbf{Projection}} &
    \begin{subfigure}{\linewidth}
        \includegraphics[width=\linewidth]{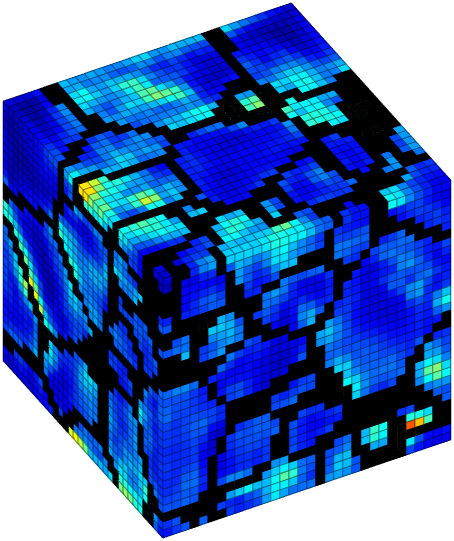}
        \caption{}        \label{fig:Fig07a-015D}
    \end{subfigure} &
    \begin{subfigure}{\linewidth}
        \includegraphics[width=\linewidth]{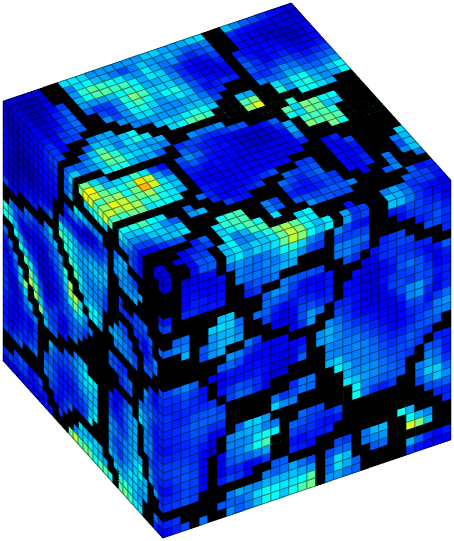}
        \caption{}        \label{fig:Fig07b-050D}
    \end{subfigure} &
    \begin{subfigure}{\linewidth}
        \includegraphics[width=\linewidth]{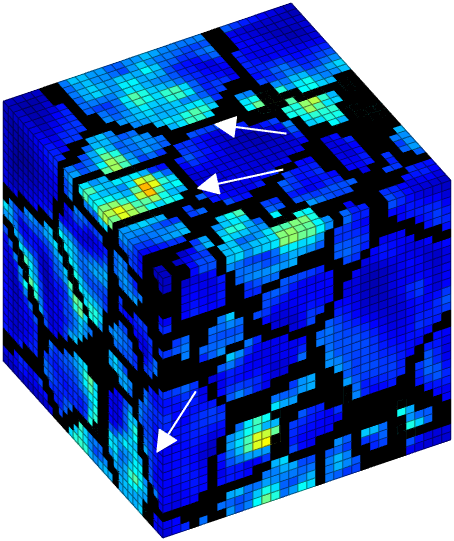}
        \caption{}        \label{fig:Fig07c-100D}
    \end{subfigure} \\

    \makecell[l]{\textbf{Active} \\ \textbf{Dislocation} \\ \textbf{Set}\\ } &
    \begin{subfigure}{\linewidth}
        \includegraphics[width=\linewidth]{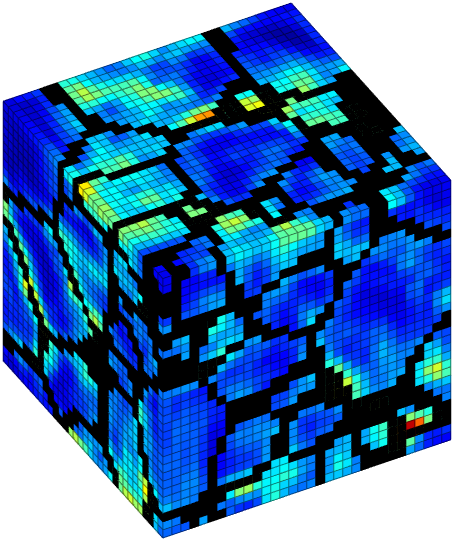}
        \caption{}        \label{fig:Fig07d-015A}
    \end{subfigure} &
    \begin{subfigure}{\linewidth}
        \includegraphics[width=\linewidth]{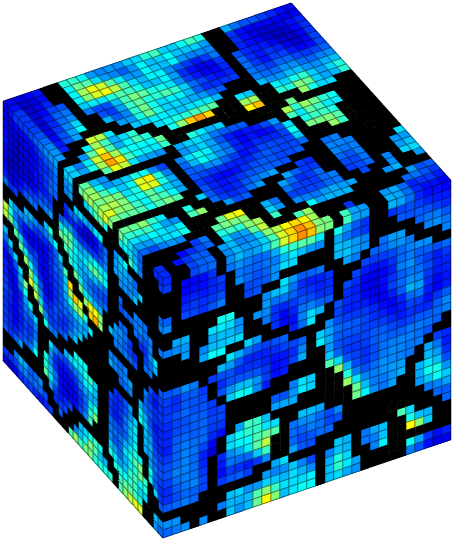}
        \caption{}        \label{fig:Fig07e-050A}
    \end{subfigure} &
    \begin{subfigure}{\linewidth}
        \includegraphics[width=\linewidth]{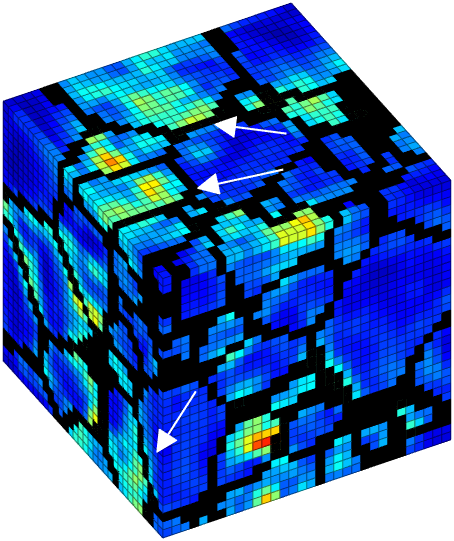}
        \caption{}        \label{fig:Fig07f-100A}
    \end{subfigure} \\

    \makecell[l]{\textbf{Shear} \\ \textbf{Approach}\\ }  &
    \begin{subfigure}{\linewidth}
        \includegraphics[width=\linewidth]{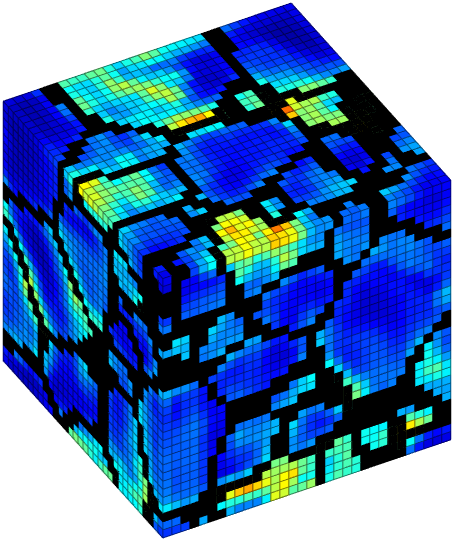}
        \caption{}        \label{fig:Fig07g-015G}
    \end{subfigure} &
    \begin{subfigure}{\linewidth}
        \includegraphics[width=\linewidth]{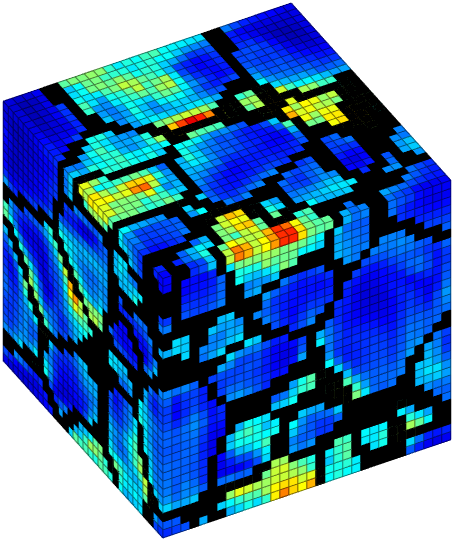}
        \caption{}        \label{fig:Fig07h-050G}
    \end{subfigure} &
    \begin{subfigure}{\linewidth}
        \includegraphics[width=\linewidth]{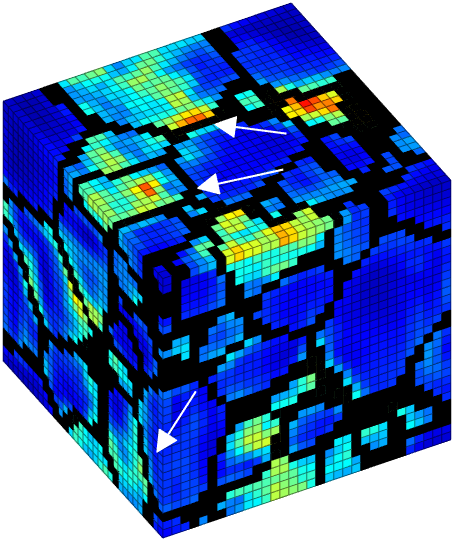}
        \caption{}        \label{fig:Fig07i-100G}
    \end{subfigure} \\

    & 
    \begin{subfigure}{\linewidth}
        \includegraphics[width=\linewidth]{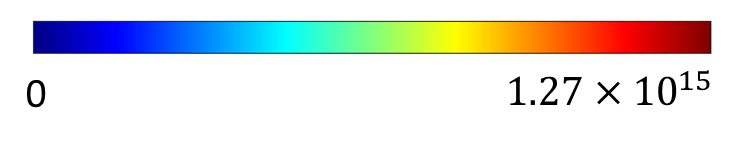}
    \end{subfigure} &
    \begin{subfigure}{\linewidth}
        \includegraphics[width=\linewidth]{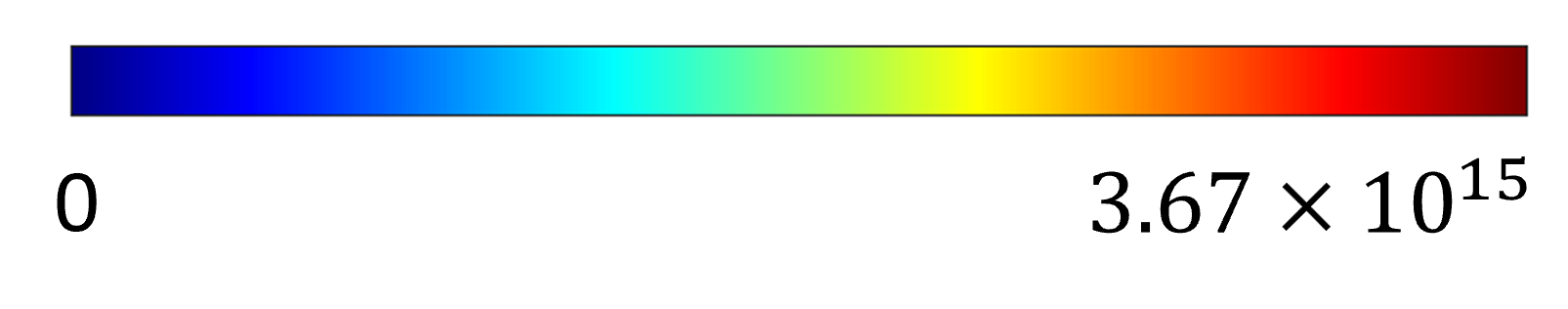}
    \end{subfigure} &
    \begin{subfigure}{\linewidth}
        \includegraphics[width=\linewidth]{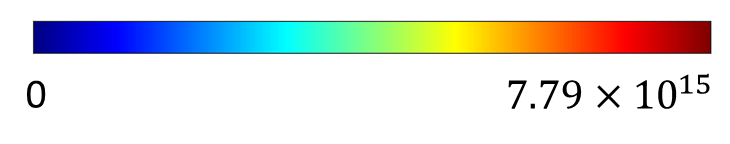}
    \end{subfigure} \\

\end{tabular}   }
\caption{GND densities as found using the three approaches, (a-c) using dislocation systems, (d-f) using the active systems, and (g-i) using the shear approach. At each strain, the figure coloring is scaled to a single colorbar. Grain boundaries are indicated in black.}
\label{fig:Fig07-gndDenNormed}
\end{figure}

The GND density plots using all three approaches at selected strains are shown in Figure~\ref{fig:Fig07-gndDenNormed}. To aid in comparing the different approaches, they are scaled such that each strain level, and thus column, shares a colorbar, from zero to the maximum GND density found at each strain. 

In general, all of the approaches showed similar patterns of GND densities in the same areas, with some examples highlighted in the 10\% strain column with white arrows, albeit with different magnitudes. Increases in strain did not significantly change the peak GND density locations. It is noted that these differences purely arise from the GND projection procedures, as the computed Nye tensors are identical for all the methods, as shown in \ref{app:Nye}.

\subsection{Comparison to predictions in Ashby's model}
A commonly used model for GND density as a function of size and strain was proposed by Ashby (1970)~\cite{Ashby1970}, listed in Eq.~\ref{eq:ashby}. It is based on the need to maintain compatibility between grains with different orientation-dependent stiffnesses. The relationship is derived from the average incompatibility (void/overlap) that would occur under deformation, a function of the average macroscopic strain, $\bar{\varepsilon}$, and the average grain diameter, $D$. Involving the Burgers vector leads to a formulation for the GND count, and dividing by the cross-sectional area gives the density.

\begin{equation} \label{eq:ashby}
    \rho_{GND} = \frac{\bar{\varepsilon}}{4bD}
\end{equation}

\subsubsection{Strain and GND density}
\label{subsubsec:GNDvStrain}

\begin{figure} \centering
    \includegraphics[width=0.7\linewidth]{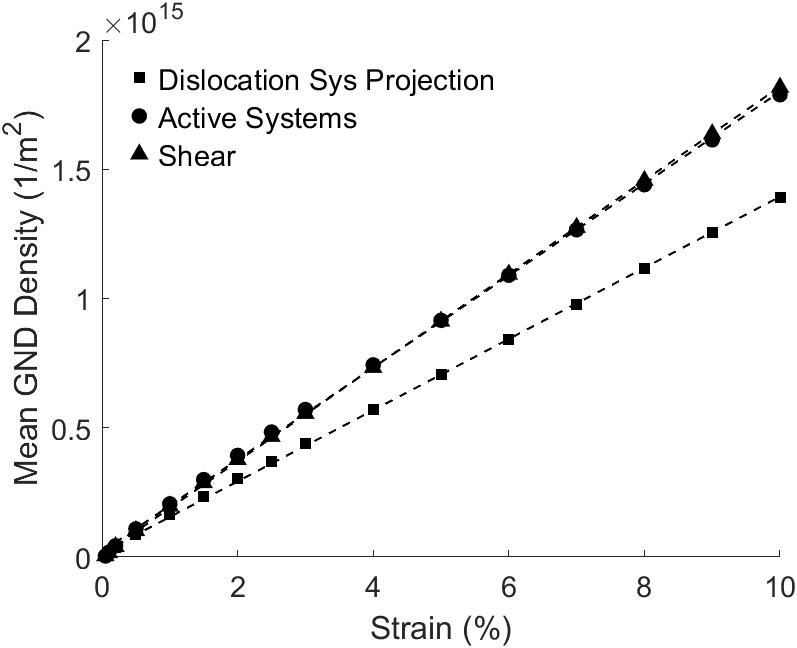}
    \caption{Average GND density vs applied strain.}
    \label{fig:Fig08-gndStrain}
\end{figure}

Based on Eq.~\ref{eq:ashby}, GND density is expected to increase linearly with strain. The models were compared to this expected trend by taking the GND density at different strains and averaging all the voxels in the RVE. 
Figure~\ref{fig:Fig08-gndStrain} shows the resulting GND density vs strain plot. Using that data, a linear fit was also found and plotted. 

From these results, the shear approach has the steepest trendline, showing the fastest increase in GND density. The differences between the dislocation approach and the slip gradient method are attributed to L2 minimization which can artificially distribute densities to both active and inactive systems~\cite{Das2018}. Restricting to active dislocation systems to address this issue (Ref~\cite{Demir2024}) increases the dislocation densities to be approximately in line with the slip gradient approach. Concluding on the best overall approach for computing polycrystal GND distributions would require a detailed comparison to experimental data.

\subsubsection{Grain size and GND density}
\label{subsubsec:GNDvSize}

To predict hardening due to GND pile-ups, the grain size effects must also be seen using the GND models. Ashby's equation captures the effect of grain size, defining the density to be linearly proportional to the inverse of the grain diameter in Eq.~\ref{eq:ashby}. To measure the models' ability to capture this trend, the ESD was taken as the grain's diameter and scaled down to 0.23 $\mu m$ and up to 2.34 $\mu m$ by adjusting the mesh sizes in postprocessing. Then GND densities were found at a strain of 10\%. 

\begin{figure}
\centering
\includegraphics[width=0.7\linewidth]{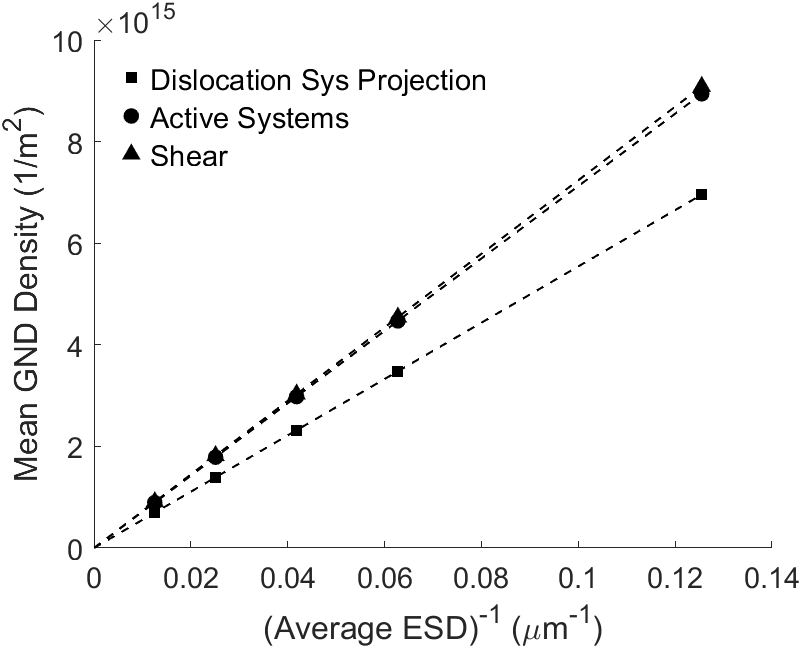}
\caption{Average GND density vs the inverse of ESD at 10\% strain.}
\label{fig:Fig09-gSizeLinear}
\end{figure}

In Figure~\ref{fig:Fig09-gSizeLinear}, the average GND density was plotted against the inverse of the average ESD at selected strains. As anticipated, a strictly linear trendline is observed. In the current formulation, the constitutive model does not include a size-dependent term, rendering the CPFE code size-independent. This introduction of gradients through the Nye tensor leads to size dependence, and linearity is observed in the GND density vs inverse ESD plots. 

\subsection{Convergence}\label{sec:convergence}

As shown in Section~\ref{subsec:scAnalysis}, all of the approaches discussed here for predicting GND densities are sensitive to the size of the elements used for meshing the RVE. This holds true for both coarse and fine discretization. Elements that are too large may underestimate gradients, by enclosing dislocations that cancel each other out. At the same time, elements that are too small require a large computational effort and may also break the continuous distribution assumptions in the Nye tensor~\cite{Arsenlis1999}. To account for this, the sample RVE (cubic, with $N$ elements per side) can be modeled with two meshes of varying densities. As long as the element side lengths are scaled in the GND calculations, the grain sizes are constant. The base ten logarithm of GND density can be plotted against the inverse of the number of elements per side, $N^{-1}$. This results in a trendline that can be extrapolated to predict other mesh sizes at a given strain level. As the number of elements increases to infinity, the trendline converges to an estimate for infinitely small elements. 

\begin{figure}
    \centering
    \includegraphics[width=0.7\linewidth]{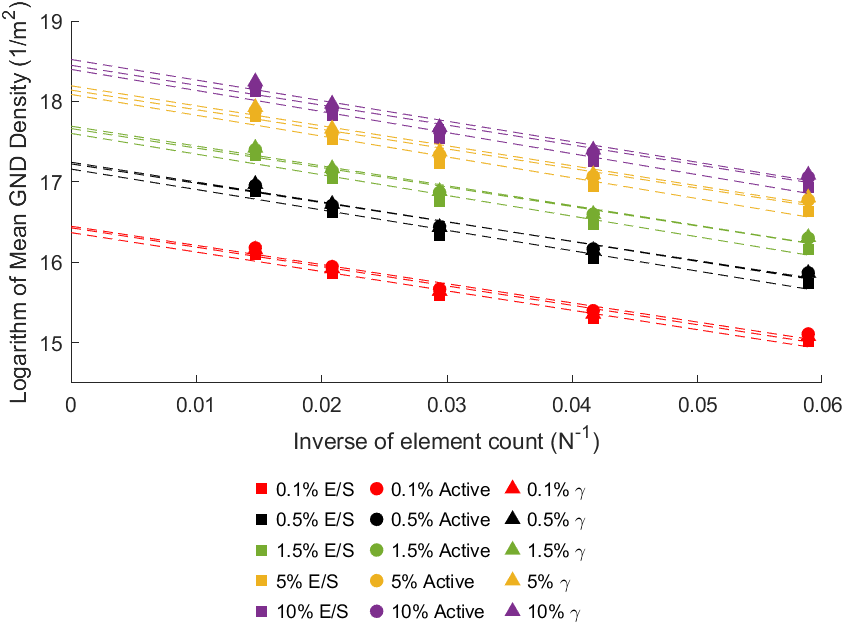}
    \caption{Base 10 logarithm of GND density vs the inverse of the number of elements per side at 0.1\%, 0.5\%, 1.5\%, 5\%, and 10\% strain. The labels  "E/S", "Active", and "$\gamma$" indicate the edge/screw dislocation system projection, the active set approach, and the shear approach, respectively.}
    \label{fig:Fig10-convergence}
\end{figure}

To show this, four additional meshes of the same RVE, with $17^3$, $24^3$, $48^3$, and $68^3$ elements, were also tested in addition to the $34^3$ mesh used for the comparisons to Ashby's equation. They were trimmed in the same way as the previous results. The resulting $\log_{10}(\rho_{GND})$ vs $N^{-1}$  plot is shown in Figure~\ref{fig:Fig10-convergence}. For any of the three models, at a given strain level, the inverse of the mesh size shows a strong linear correlation with the logarithm of the average GND density. Consequently, at any given strain level, using two meshes with different mesh sizes is enough to predict the GND density for any mesh size. Importantly, it must be noted that the mesh sensitivity is strain dependent. Thus, this process needs to be repeated at every strain level of interest.

\section{GND Accumulation at Microstructural Heterogeneities}
The GND models were then applied to study areas of research for GNDs, such as investigating the relationships of GNDs at grain boundaries, multi-grain junctions, and grain interiors~\cite{Lee2025,Clair2011,Huang2024}. As such, they were used in studies on GND densities at grain junctions, as well as the effect that the Schmid Factor has on GND distributions.

\subsection{Multigrain junctions} \label{subsec:multigrainJuns}

Grain boundaries, particularly multigrain junctions (those involving three or more grains), are known to exhibit increased stress~\cite{Lee2025}. Furthermore, multiple grains with mismatches in slip transmission modes are known to correlate with higher GND densities~\cite{Clair2011,Huang2024}. To compare how the models capture this effect, the GND density was estimated by all three approaches and compared in three regions: multigrain junctions (MG), grain boundaries involving only two grains (GB), and grain interior voxels (GI). 

All the models capture increases in GND density within a single category (ex: grain interior or grain boundaries) as the strain increases. Secondly, as expected from experimental work~\cite{zhong2024}, both the multigrain junctions and GB resulted in elevated GND densities relative to the rest of the model, with multigrain junctions always resulting in the highest average density. 

\begin{figure}
    \centering
    \includegraphics[width=0.7\linewidth]{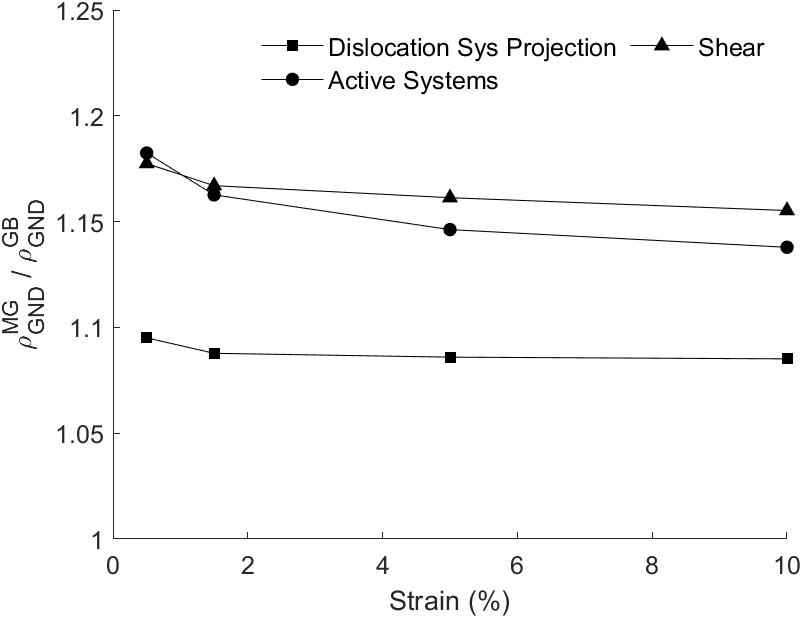}
    \caption{The ratio of the average GND density in the multigrain junctions to boundaries between two grains.}
    \label{fig:Fig11-TJvGB}
\end{figure}

The consistency of the latter trend is seen in Figure~\ref {fig:Fig11-TJvGB}, which shows the ratio of the average GND density in multigrain junctions ($\rho_{GND}^{MG}$) to that in grain boundaries ($\rho_{GND}^{GB}$)  at different strain levels. The ratio $\rho_{GND}^{MG}/\rho_{GND}^{GB}$ is nearly constant up to 10\% strain for the projection approach, with the density in multigrain junctions about 9\% greater than the density in grain boundaries. For the active system and shear approaches, multigrain junctions initially see about 18\% higher average densities, but demonstrate a decreasing ratio with strain, ending around 14-15\% higher.

\begin{figure}
    \centering
    \includegraphics[width=0.7\linewidth]{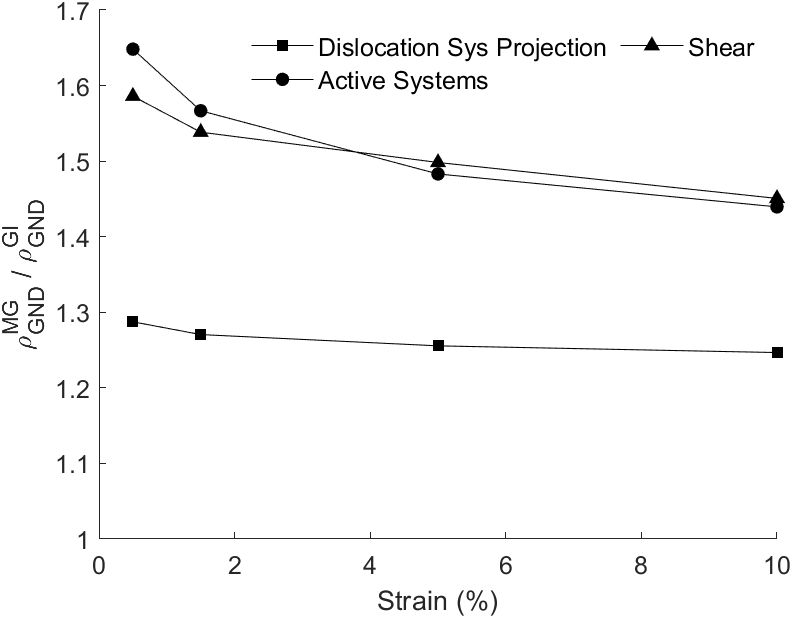}
    \caption{The ratio of the average GND density in the multigrain junctions to grain interiors.}
    \label{fig:Fig12-TJvGI}
\end{figure}

Figure~\ref{fig:Fig12-TJvGI} shows the ratio of GNDs between multigrain junctions ($\rho_{GND}^{MG}$) and grain interiors($\rho_{GND}^{GI}$). Compared to the trend observed for the GND density ratio between multigrain junctions and grain boundaries, all three methods show larger initial values and greater decreases in the ratio $\rho_{GND}^{MG}/\rho_{GND}^{GI}$ as the strain increases. The projection approach initially has around 29\%  more dislocations at multigrain junctions than in the grain interiors, dropping by 4\% by the end of the simulation. The active system falls from 64.8\%-43.9\% and the shear approach from 58.6\% to 45\%. This decrease in the ratio of GND densities between the grain boundaries and the interior of the grain with increased strain matches experimental observations, such as in Ref.~\cite{song2024}. 

The drop is observed because at low strains, GNDs primarily accumulate near grain boundaries to accommodate incompatibilities between neighboring grains. As strain rises, dislocations intensify throughout the grain, including in the interior, while the boundary regions have slower accumulation due to limited space, leading to annihilation and local reorientation~\cite{yoo2022correlating}.

\subsection{GND accumulation at grain boundaries}
A variety of different parameters have been used in literature to model slip transmissions and pileup at grain boundaries, including the Luster-Morris factor~\cite{Luster1995} and LRB factor~\cite{LRB1989}. Others combine these with the Schmid factor to incorporate slip system activity~\cite{Bieler2014}. Here, the Schmid factor is chosen as one possible indicator for GND accumulation at grain boundaries because the interface between high and low Schmid factor grains would likely show high GND accumulation, which is confirmed from the examples.

\begin{figure}
\centering
\scalebox{0.8}{
\begin{tabular}{
    >{\centering\arraybackslash}m{0.43\textwidth}
    >{\centering\arraybackslash}m{0.18\textwidth}
    >{\centering\arraybackslash}m{0.36\textwidth}}
    
    \begin{subfigure}{\linewidth}
        \includegraphics[width=\linewidth]{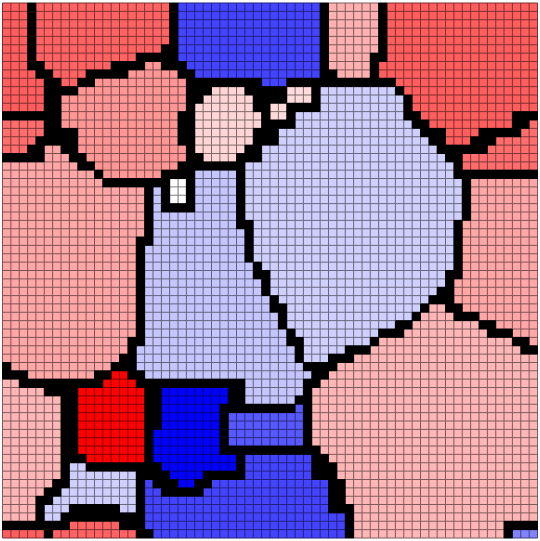}
        \caption{}
    \end{subfigure} & 
    \begin{subfigure}{\linewidth}
         \includegraphics[height=2.5\linewidth, keepaspectratio]{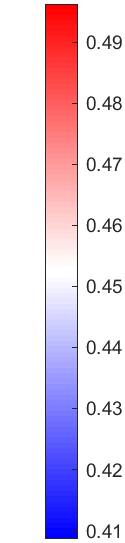}
    \end{subfigure} &
     \begin{subfigure}{\linewidth} 
        \includegraphics[width=\linewidth]{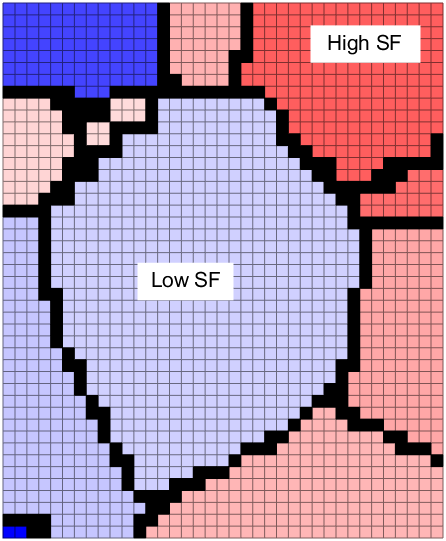}
        \caption{} \label{fig:Fig13c-sfMapSubSec}
    \end{subfigure} \\
\end{tabular} }
\caption{The (a) full cross-section of the sample microstructure and (b) subsection used in this study, colored by the Schmid factor with the grain boundaries indicated in black.}
\label{fig:Fig13-sfMap}
\end{figure}

To select a region for observation, the Schmid factor was found from the undeformed orientations of the grains. A cross-section of the microstructure, colored by the SF, is shown in Figure~\ref{fig:Fig13-sfMap}. Each voxel is split into eight elements, each colored by the same value, to show better detail near grain boundaries, which are indicated in black. The subregion of the microstructure shown in Fig.~\ref{fig:Fig13c-sfMapSubSec} has highlighted grains with a high Schmid factor and a comparatively low Schmid factor. 

\begin{figure}
\centering
\scalebox{1}{
\begin{tabular}{
    >{\centering\arraybackslash}m{0.29\textwidth}
    >{\centering\arraybackslash}m{0.29\textwidth}
    >{\centering\arraybackslash}m{0.29\textwidth}
    >{\centering\arraybackslash}m{0.11\textwidth}}
    
    \begin{subfigure}{\linewidth}
        \includegraphics[width=\linewidth]{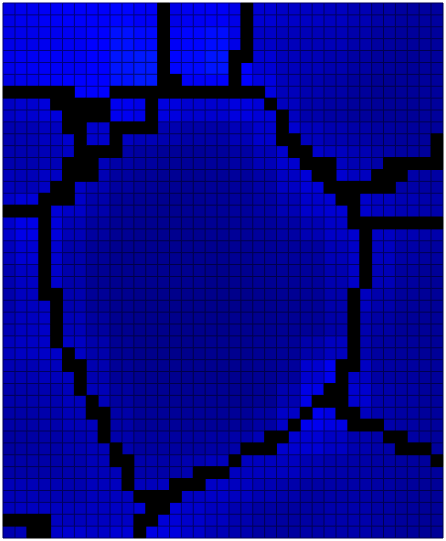}
        \caption{}
    \end{subfigure} & 
    \begin{subfigure}{\linewidth}
        \includegraphics[width=\linewidth]{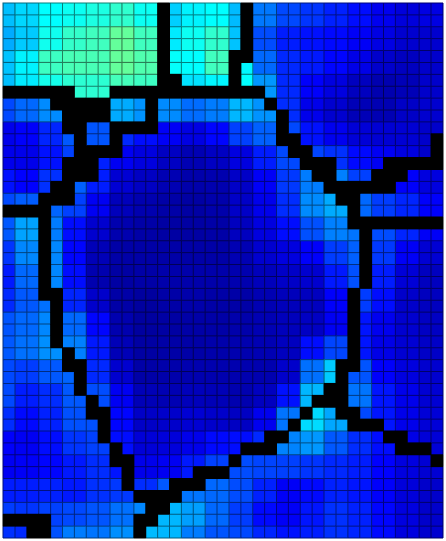}
        \caption{}
    \end{subfigure} &
    \begin{subfigure}{\linewidth}
        \includegraphics[width=\linewidth]{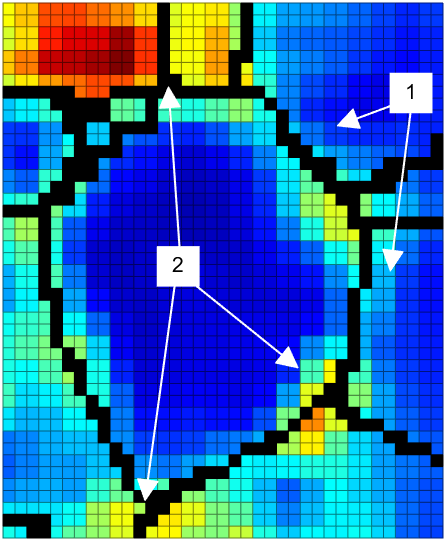}
        \caption{} \label{fig:Fig14c-gndShearGB100Label}
    \end{subfigure} &
    \begin{subfigure}{\linewidth} 
        \raisebox{3ex}{\includegraphics[height=3.2\linewidth]{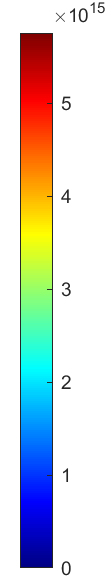}}
    \end{subfigure} \\
\end{tabular} }
\caption{GND densities found using the shear approach, colored by a single colorbar, at (a) 1.5\%, (b) 5\%, and (c) 10\% strain. Grain boundaries are indicated in black. Areas labeled `1' show accumulation along a high SF/ low SF boundary, and `2' show accumulation at a multigrain junction.}
\label{fig:Fig14-gndShearGBLay}
\end{figure}

The shear approach is used for computing the GND densities at 1.5\%, 5\%, and 10\% in Fig.~\ref{fig:Fig14-gndShearGBLay}. At all three strain levels, elevated GND densities are observed at grain boundaries, especially at junctions with high SF and low SF, as shown in Fig.~\ref{fig:Fig14c-gndShearGB100Label} with the label `1'. Furthermore, particularly high densities are observed by multigrain junctions, with a few examples labeled `2'. Other effects may also lead to areas of higher densities, such as grain size effects and other grain boundaries not observed in the two-dimensional cross-section.

\begin{figure}
\centering
\scalebox{0.89}{
\begin{tabular}{
    >{\centering\arraybackslash}m{0.10\textwidth}
    >{\centering\arraybackslash}m{0.30\textwidth}
    >{\centering\arraybackslash}m{0.12\textwidth}
    >{\centering\arraybackslash}m{0.30\textwidth}
    >{\centering\arraybackslash}m{0.12\textwidth}
}

    \textbf{1.5\%} & \begin{subfigure}{\linewidth}
        \includegraphics[width=\linewidth]{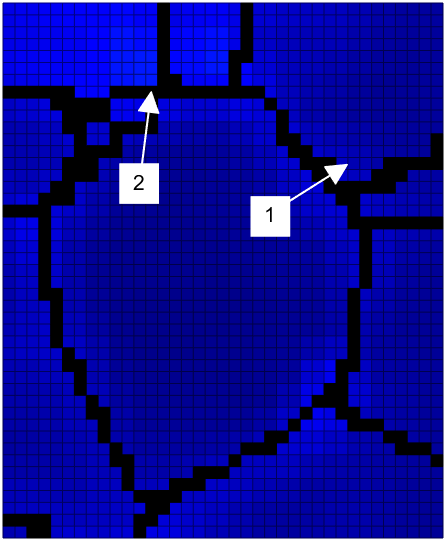}
        \caption{} \label{fig:Fig15a}
    \end{subfigure} & 
    \begin{subfigure}{\linewidth}
        \raisebox{4ex}{\includegraphics[height=3.2\linewidth,keepaspectratio]{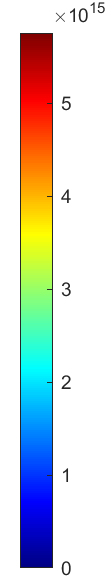}}
    \end{subfigure} &
    \begin{subfigure}{\linewidth}
        \includegraphics[width=\linewidth]{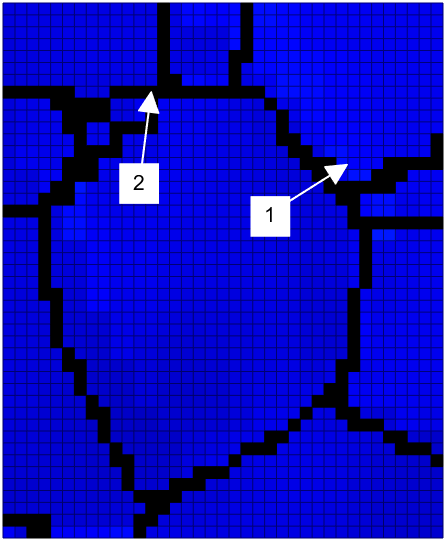}
        \caption{} \label{fig:Fig15b}
    \end{subfigure} &
    \begin{subfigure}{\linewidth}
        \raisebox{2ex}{\includegraphics[height=3\linewidth,keepaspectratio]{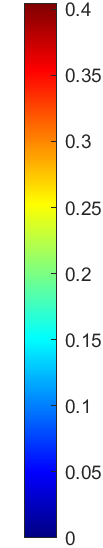}}
    \end{subfigure} \\

    \textbf{5\%} & \begin{subfigure}{\linewidth}
        \includegraphics[width=\linewidth]{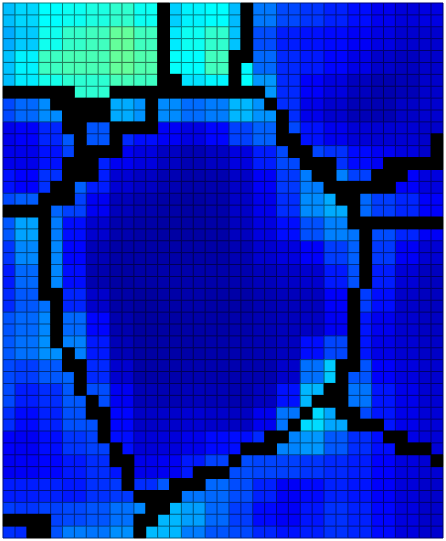}
        \caption{} \label{fig:Fig15c}
    \end{subfigure} & 
    \begin{subfigure}{\linewidth}
        \raisebox{2ex}{\includegraphics[height=3\linewidth,keepaspectratio]{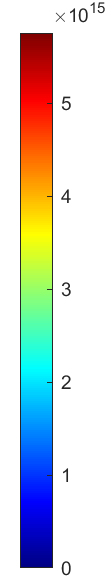}}
    \end{subfigure} &
    \begin{subfigure}{\linewidth}
        \includegraphics[width=\linewidth]{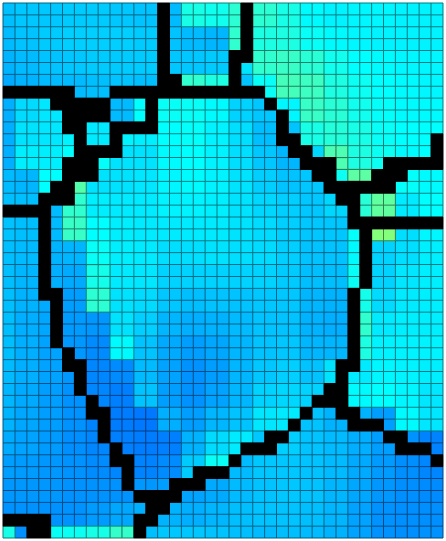}
        \caption{} \label{fig:Fig15d}
    \end{subfigure} &
    \begin{subfigure}{\linewidth}
        \raisebox{4ex}{\includegraphics[height=3.2\linewidth,keepaspectratio]{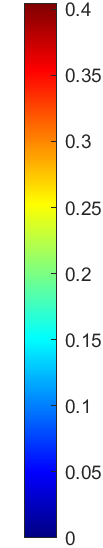}}
    \end{subfigure} \\

    \textbf{10\%} & \begin{subfigure}{\linewidth}
        \includegraphics[width=\linewidth]{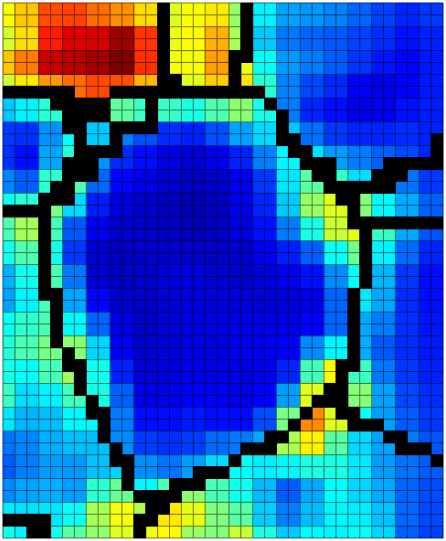}
        \caption{} \label{fig:Fig15e}
    \end{subfigure} & 
    \begin{subfigure}{\linewidth}
        \raisebox{2ex}{\includegraphics[height=3\linewidth,keepaspectratio]{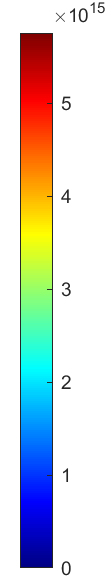}}
    \end{subfigure} &
    \begin{subfigure}{\linewidth}
        \includegraphics[width=\linewidth]{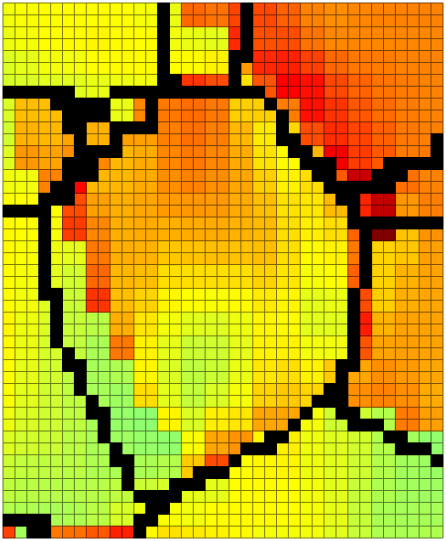}
        \caption{} \label{fig:Fig15f}
    \end{subfigure} &
    \begin{subfigure}{\linewidth}
        \raisebox{4ex}{\includegraphics[height=3.2\linewidth,keepaspectratio]{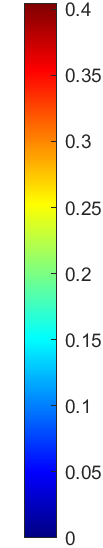}}
    \end{subfigure} \\

\end{tabular} }
\caption{GND densities found using the shear approach (left) and sum of shears on all slip systems (right) at 1.5\% (a,b), 5\% (c,d), and 10\% (e,f) strain. Grain boundaries are indicated in black. Regions of high total shear do not necessarily correlate with high GND density computed using the shear (`1') and vice versa (`2').}
\label{fig:Fig15-gndVsShear}
\end{figure}

It is important to note that areas of high shear do not directly correlate with high GND density. The formulation relies on the gradient of shear (Eq.~\ref{eq:shearFinal}), rather than the total shear. This effect is shown in Fig.~\ref{fig:Fig15-gndVsShear}. The GND density using the shear approach is on the right, and the sum of shear is shown in the left column.

\section{Conclusion}

In this study, three approaches were used to determine total GND densities in samples with small deformations. These included the use of $\text{curl}(\mathbf{F}^p)$ to find Nye's tensor before (1) decomposing it onto the edge and screw dislocation systems using L2 minimization, (2) decomposing it onto dislocation systems associated with shearing slip systems, and (3) using the shear to find GND densities on the slip systems. Single crystal simulations were used to compare results to known analytical models. Then, in a series of studies on a sample FCC microstructure, trends in GND density estimates using the three models were compared to those predicted by Ashby's equation and compared to one another. The GND models were used to observe the relationship between the GND density at multigrain junctions, two-grain boundaries, and the interior of grains. Finally, the effects of the boundaries and the SF were explored. This work contributes to GND literature by identifying methodological discrepancies in GND quantification across different techniques and practical implications for polycrystal modeling.

From this work, there are five primary conclusions:

\begin{enumerate}
    \item The slip gradient approach shows the greatest similarity to analytically predicted results in single crystal cases by both showing zero GND density under symmetric loading and predicting GND densities closer to analytically predicted ones than those from the dislocation systems across a range of mesh discretizations. In polycrystal cases, the dislocation projection approach shows lower GNDs than slip gradient methods. With the selection of the active dislocation set, the differences are significantly reduced. Concluding on the best approach, however, would require future comparison to experimental data.
    \item All of the techniques show similar distributions of GND densities with varying magnitudes. All three approaches follow Ashby's relation, showing increased GND density with (i) an increase in strain and (ii) a decrease in grain size, allowing their use in a qualitative study of trends observed in experiments. 
    \item To determine the GND results for an RVE with any level of mesh refinement, two samples can be used to extrapolate a linear model at any given strain level. The mesh size dependence means that calculations reliant on GND density estimates must be calibrated or designed to be mesh-size agnostic to avoid propagating an error.
    \item All the models capture the increase in the GNDs observed near multigrain junctions and grain boundaries compared to the grain interior. Multigrain junctions average about a 10-18\% increase over basic grain boundaries and about 25-65\% greater than the grain interiors, depending on the approach and the strain. The latter ratio is particularly strain dependent, decreasing with strain due to faster buildup of GNDs in the grain interior, an aspect captured by all the methods.
    \item Elevated GND densities are observed near the grain boundaries of low and high Schmid factor grains, accumulating predominantly within high SF grains, demonstrating incompatibility-driven accumulation of GNDs.
\end{enumerate}

\section*{Data Availability}
Data will be made available upon publication.

\section*{Acknowledgments and Funding}
This work is funded by the U.S. Department of Energy, Office of Basic Energy Sciences, Division of Materials Sciences and Engineering under Award \#DE-SC0008637. The work was supported in part through computational resources and services provided by Advanced Research Computing at the University of Michigan, Ann Arbor and Bridges-2 at Pittsburgh Supercomputing Center through allocation MSS160003 from the Advanced Cyberinfrastructure Coordination Ecosystem: Services \& Support (ACCESS) program, which is supported by National Science Foundation grants \#2138259, \#2138286, \#2138307, \#2137603, and \#2138296. The authors would also like to thank Prof.~John Allison (Materials Science and Engineering) at the University of Michigan for his input during the development of this work. Furthermore, the authors would like to thank Prof.~Wolfgang Pantleon (Materials and Surface Engineering) at the Technical University of Denmark for useful insights.

\appendix
\section{Nye tensor comparison}
\label{app:Nye}
The intermediate calculation in the GND density approaches, finding the Nye tensor, must be equivalent per Eq.~\ref{eq:nyeShear}, whether computed from the curl of the plastic deformation tensor or the shear. Given equal results for the tensor's components, any differences in GND densities arise from further calculations.

\begin{figure}[h]
\centering
\scalebox{0.83}{
\begin{tabular}{
    >{\arraybackslash}m{0.16\textwidth}
    >{\centering\arraybackslash}m{0.27\textwidth}
    >{\centering\arraybackslash}m{0.27\textwidth}
    >{\centering\arraybackslash}m{0.27\textwidth}
}

    \textbf{Strain:} & \textbf{1.5\%} & \textbf{5\%} & \textbf{10\%} \\ 

   \makecell[l]{\textbf{Direct} \\ $\mathrm{curl}(\mathbf{F}^p)$ \\ } &
    \begin{subfigure}{\linewidth}
        \includegraphics[width=\linewidth]{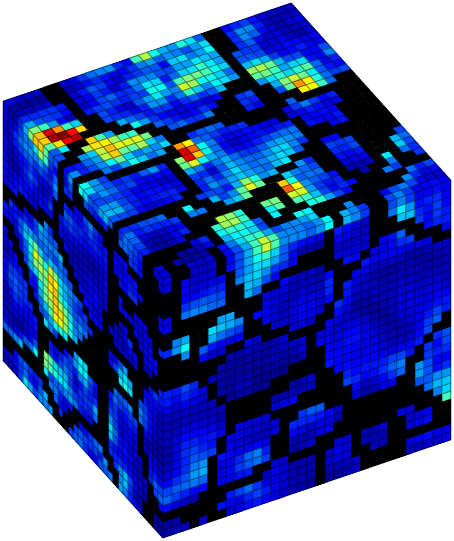}
        \caption{} \label{fig:Fig16a-015S}
    \end{subfigure} &
    \begin{subfigure}{\linewidth}
        \includegraphics[width=\linewidth]{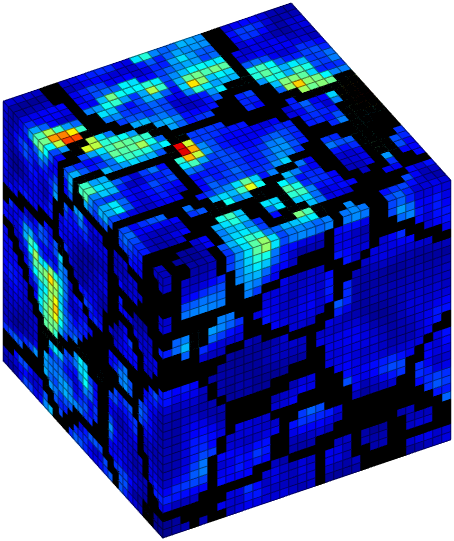}
        \caption{} \label{fig:Fig16b-050S}
    \end{subfigure} &
    \begin{subfigure}{\linewidth}
        \includegraphics[width=\linewidth]{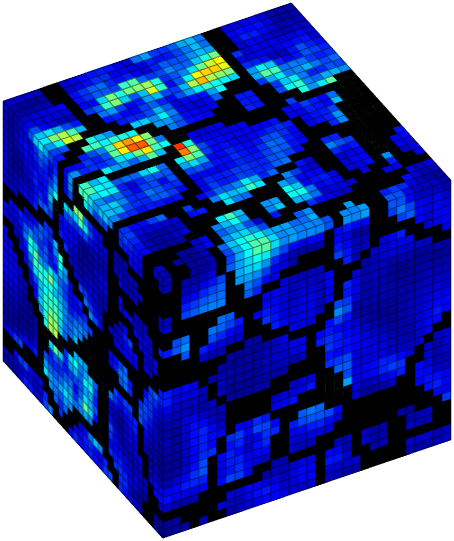}
        \caption{} \label{fig:Fig16c-100S}
    \end{subfigure} \\

    \makecell[l]{\textbf{} \\ \textbf{Shear}\\ \textbf{}} &
    \begin{subfigure}{\linewidth}
        \includegraphics[width=\linewidth]{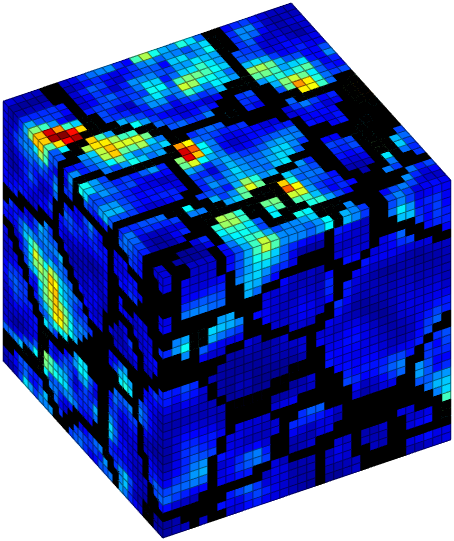}
        \caption{} \label{fig:Fig16d-015G}
    \end{subfigure} &
    \begin{subfigure}{\linewidth}
        \includegraphics[width=\linewidth]{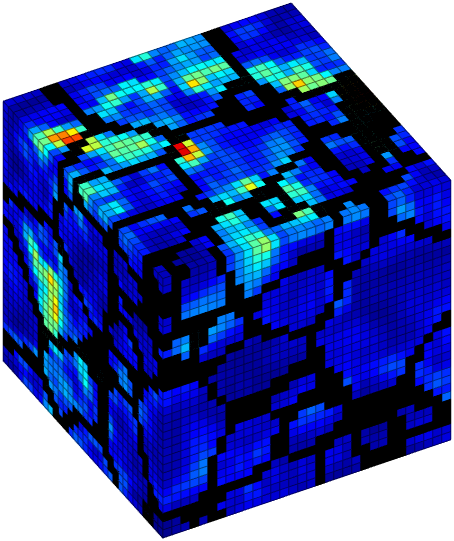}
        \caption{} \label{fig:Fig16e-050G}
    \end{subfigure} &
    \begin{subfigure}{\linewidth}
        \includegraphics[width=\linewidth]{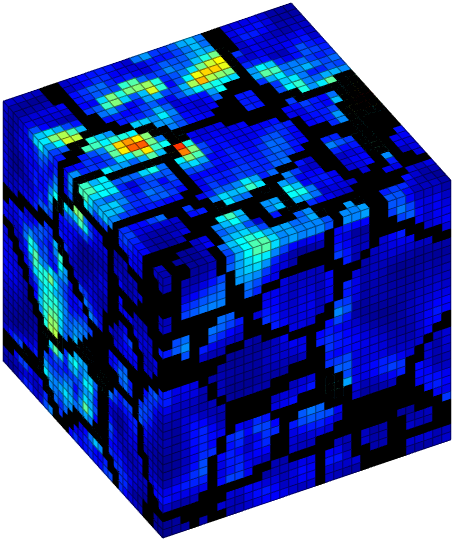}
        \caption{} \label{fig:Fig16f-100G}
    \end{subfigure} \\

    & 
    \begin{subfigure}{\linewidth}
        \includegraphics[width=\linewidth]{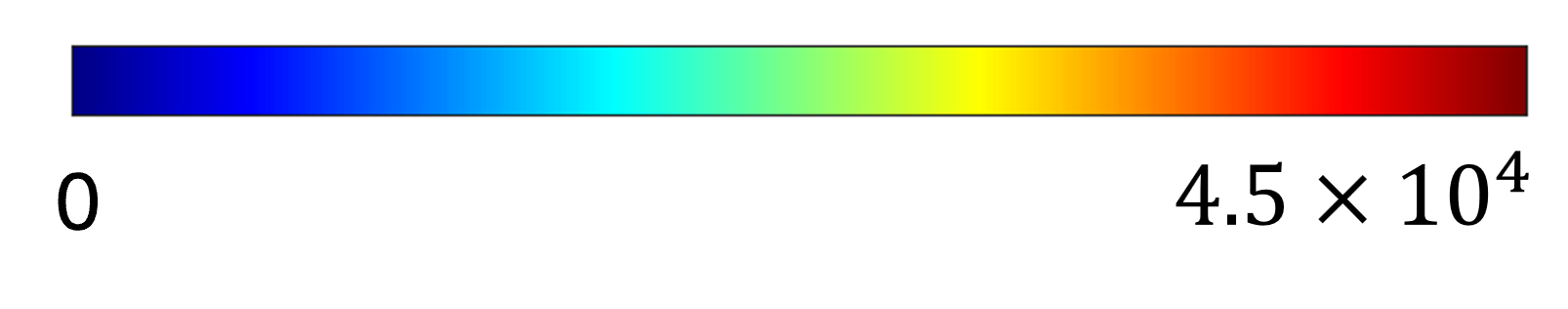}
    \end{subfigure} &
    \begin{subfigure}{\linewidth}
        \includegraphics[width=\linewidth]{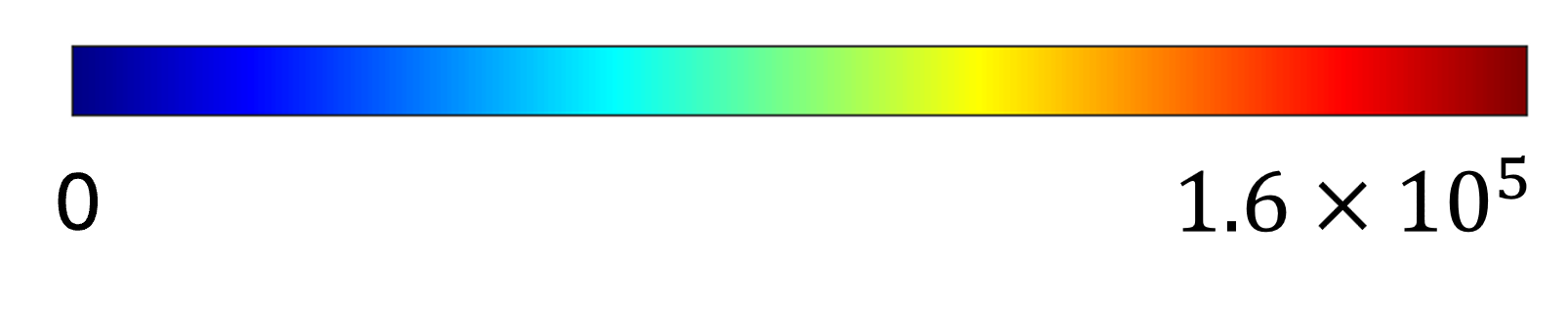}
    \end{subfigure} &
    \begin{subfigure}{\linewidth}
        \includegraphics[width=\linewidth]{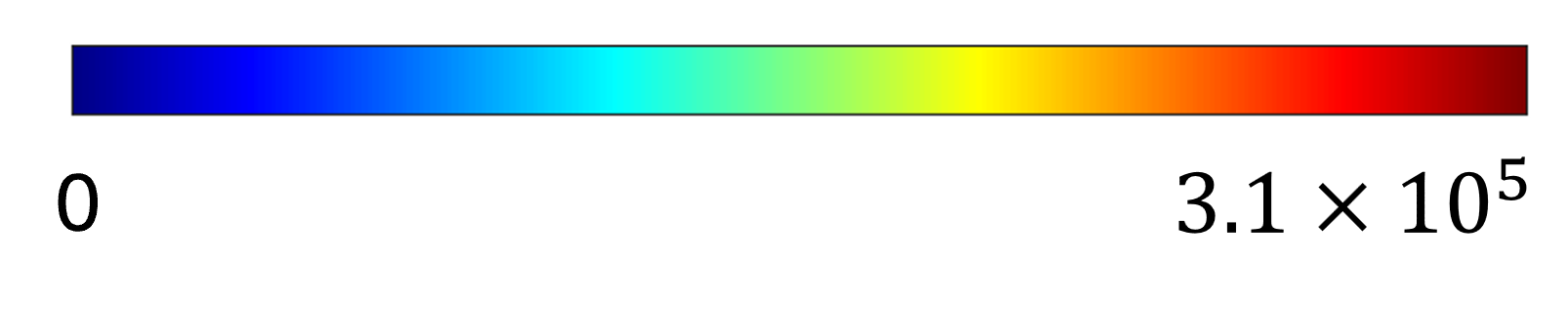}
    \end{subfigure} \\

\end{tabular}   }
\caption{The $G_{33}$ component of the Nye tensor, at 1.5\%, 5\%, and 10\% strain, as found (a-c) directly from $\text{curl}(\mathbf{F}^p)$ and (d-f) from the shear method. At each strain, the figure coloring is scaled to a single colorbar. Grain boundaries are indicated in black.}
\label{fig:Fig16-nyeTenNormed}
\end{figure} 

The $\mathbf{G}_{33}$ component of the Nye tensor at 1.5\%, 5\%, and 10\% strain is shown in Figure~\ref{fig:Fig16-nyeTenNormed}. Two methods for finding the Nye tensor are shown - the first, using $\text{curl}(\mathbf{F}^p)$ (\ref{fig:Fig16a-015S}-\ref{fig:Fig16c-100S}), which is shared by the first two approaches, and the other using shear (\ref{fig:Fig16d-015G}-\ref{fig:Fig16f-100G}).

To draw comparisons between the two approaches to finding the Nye tensor, the plots are shown with a single colorbar at each strain level. The general distribution is closely matched between the two approaches. Furthermore, at all the observed strains, even beyond those shown in the figure, the maximum values of the Nye tensor maintain this similarity. The maximum value of the $\mathbf{G}_{33}$ component of the Nye tensor at different strain levels is listed in Table~\ref{tab:nyeVals}, showing a maximum less than 1.6\% difference. Thus, as anticipated, the Nye tensors are in agreement, as seen in Fig.~\ref{fig:Fig16-nyeTenNormed}, and differences in GND density distributions between the projection and shear approaches are the result of the subsequent decomposition after the Nye tensor is found.

\begin{table}[h]
\centering
\caption{Maximum values of the Nye tensor using the curl of the deformation gradient and shear approaches at selected strains.}
\label{tab:nyeVals}
\begin{tabular}{lccccccc} \hline
Strain & 0.2\% & 0.5\% & 1\% & 1.5\% & 2\% & 5\% & 10\% \\ \hline
$\text{curl}(\mathbf{F}^p)$ & 7370 & 16076 & 29640 & 44878 & 59724 & 155999 & 309621 \\
Shear & 7365 & 16066 & 29545 & 44642 & 59269 & 155732 & 304848 \\ 
\% Difference & 0.076 & 0.064 & 0.322 & 0.531 & 0.768 & 0.171 & 1.566 \\ \hline
\end{tabular}
\end{table}

\bibliographystyle{elsarticle-num} 
\bibliography{references.bib}

\end{document}